\documentclass[aps,pra,twocolumn,groupedaddress]{revtex4-2}

\usepackage{amsmath}
\usepackage{braket}
\usepackage{amssymb}
\usepackage{graphicx}
\usepackage[caption=false]{subfig}
\usepackage[breaklinks=true]{hyperref}
\usepackage{breakcites}

\begin{document}


\title{Orbital Variational Adiabatic Hyperspherical Method Applied to Bose-Einstein Condensates}




\author{Hyunwoo Lee}
\email[]{lee2232@purdue.edu}
\affiliation{Department of Physics and Astronomy, Purdue University, West Lafayette, IN 47907 USA}
\author{Chris H. Greene}
\email[]{chgreene@purdue.edu}
\affiliation{Department of Physics and Astronomy, Purdue University, West Lafayette, IN 47907 USA}
\affiliation{Purdue Quantum Science and Engineering Institute,
Purdue University, West Lafayette, IN 47907 USA}


\date{December 21, 2020}

\begin{abstract}
A variational basis set motivated by mean-field theory is utilized to describe the Bose-Einstein condensate within the adiabatic hyperspherical coordinate framework.  The simplest single-orbital variant of this treatment reproduces many of the ground state properties predicted by the Gross-Pitaevskii equation.  But a multi-orbital improvement to the basis set yields a better representation of particle correlations and of the critical number where the condensate collapses for a negative two-body scattering length.  The method also produces systematic deviations from Bogoliubov theory for the fundamental monopole excitation frequency.

\end{abstract}


\maketitle


\section{Introduction}

Bose-Einstein condensates provide a deep and insightful laboratory for understanding the properties of quantum many-body systems \cite{RevModPhys.80.885}. The usual theoretical method used to describe dilute atomic systems  writes a mean-field product wavefunction approximation that generates a nonlinear term which accounts for the mutual low-energy s-wave interactions of the atoms. However, as far as it is assumed valid, the underlying many-particle Schr{\"o}dinger equation is a \textit{linear} equation, and it brings into question whether a direct treatment of the many-body Hamiltonian can be solved sufficiently accurately to demonstrate consistency with observations of nonlinear physics, such as bright and dark solitons. There have been several different attempts to directly treat the many-body Hamiltonian, most prominently Monte Carlo calculations \cite{PhysRevLett.77.3695, PhysRevA.73.063619}. This paper applies instead the adiabatic hyperspherical methods, which have been widely utilized with success in describing few-body systems. Of many examples, perhaps the greatest triumphs are the early theoretical prediction \cite{EFIMOV1973157} and experimental confirmation \cite{Kraemer_2006_nature} decades later of the Efimov spectrum of three resonantly interacting bosons. A serious difficulty, however, is in being able to compute the adiabatic potential curves in the first place, made exponentially harder as more particles are considered.

The earliest treatment of the bosonic many-body problem within the hyperspherical framework is given by Ref. \cite{PhysRevA.58.584}. Here, the simplifying assumption (called the K-Harmonic approximation) is that the hyperangular behavior of the interacting $N$-boson system in its ground state is constant, which corresponds to the lowest hyperspherical harmonic. This leads to a radial Schr{\"o}dinger equation in a single adiabatic coordinate, the hyperradius $R$, which gives an intuitive picture of the energy dependence of the atomic cloud on its root-mean-square cloud radius. Once the adiabatic potential is known, it gives estimates of the ground-state energy and monopole breathing mode frequencies, both of which show interesting differences from the numerical mean-field results. The method also predicts the critical particle number $N_c$ for collapse of an attractive condensate with two-body scattering length $a_s < 0$ to be given by $N_c \frac{|a_s|}{l_t} \approx 0.671$, where $l_t = \sqrt{\frac{\hbar}{m \omega}}$ is the trap length scale. In comparison, a variational treatment of the Gross-Pitaevskii (GP) equation using a gaussian ansatz results in an energy functional in terms of the gaussian width \cite{PhysRevA.56.1424}, which actually looks remarkably close to the K-Harmonic adiabatic potential $U(R)$. Hence, the two methods give nearly identical predictions for $N_c$; for reference, numerical solution of the GP equation \cite{PhysRevA.64.055602} gives $N_c \frac{|a_s|}{l_t} \approx 0.575$. The K-Harmonic approximation was generalized to treat anisotropic traps \cite{Kim_1999, PhysRevA.70.063617} and was also applied to a degenerate Fermi gas using a Slater determinantal trial wavefunction composed of trap eigenstates \cite{PhysRevA.74.053624, rittenhouse_jpchem2009}.

Of course, though qualitative insights can be gleaned from the simple analytic results, the K-Harmonic approximation is both restrictive and over-simplifying, so various approaches have been formulated to better treat the many-body Hamiltonian. One approach is to apply a Faddeev decomposition for the adiabatic channel function and solve an integro-differential equation \cite{PhysRevA.65.051601, PhysRevA.66.032507}. Another is a diffusive Monte Carlo calculation \cite{blume_jchemp2000} of the lowest eigenvalue for the adiabatic Hamiltonian (of fixed $R$) with modest $N$. Yet another attempt is a potential harmonics expansion method \cite{PhysRevA.70.063601, Chakrabarti_2005}, which includes more than one hyperspherical harmonics that effectively only accounts for two-body correlations. And, finally, one may assume a Jastrow-type ansatz for the channel function and apply Bethe-Peierls boundary conditions \cite{PhysRevA.97.033608}.

This paper presents an alternative variational method for computing the lowest adiabatic hyperspherical potential $U(R)$ of a spherically symmetric Bose-Einstein condensate. The many-body wavefunction is assumed to be given by a symmetric product of a chosen orbital $\phi( \vec{r_i} )$, or by linear combinations of such direct products, where $\phi$ might or might not be the mean-field orbital. While this ansatz holds no information whatsoever regarding the behavior of the system with respect to inter-particle distances, in a way that the methods of Refs. \cite{PhysRevA.65.051601} and \cite{PhysRevA.97.033608} would have, there are several clear advantages. The numerical method works well for large and easily variable effective particle number $N_0$ used to find the basis orbital from the GP equation, which does not need to coincide with the true particle number $N$ being treated in the Hamiltonian. Depending on the choice of the orbital, a large family of potential curves $U(R)$ that varies just on a few parameters can now be obtained to great intuitive use. Coupling between different potential curves can be computed, and a limited diagonalization of the adiabatic Hamiltonian is possible. Finally, the method is a direct, straightforward generalization of the K-Harmonic approximation and serves as an interconnecting bridge between the adiabatic hyperspherical formalism and the usual mean-field approach.

\section{Methods}

\subsection{Basic Formalism}

Consider the following many-body Hamiltonian for $N$ spin-less (or spin-polarized) bosons in a spherical trap. Represent the mutual two-body interactions by an s-wave Fermi pseudopotential \cite{PhysRevA.65.043613}, where the low-energy scattering length $a_s$ describes the shape of the two-body wavefunction outside the range of actual interaction potential, while the detailed shape of the short-range potential is regarded as irrelevant. $g = \frac{4 \pi \hbar^2 a_s}{m}$ is the effective interaction strength.

\begin{equation}
    H = - \frac{\hbar^2}{2m} \displaystyle\sum_{i=1}^N \nabla_i^2 + \frac{1}{2} m \omega^2 \displaystyle\sum_{i=1}^N r_i^2 + g \displaystyle\sum_{i<j} \delta ( \vec{r_i} - \vec{r_j} )
\end{equation}

Recall \cite{RevModPhys.71.463, RevModPhys.73.307} that the most commonly utilized approach is to consider a simple exchange-symmetric many-body wavefunction of the form $ \Psi = \displaystyle\prod_{i=1}^N \phi ( \vec{r_i} ) $ and variationally minimize the energy $E[\phi]$, which leads to the number-conserving form of the Gross-Pitaevskii (GP) equation:

\begin{equation}\label{GPeqn}
    - \frac{\hbar^2}{2m} \nabla^2 \phi + \frac{1}{2} m \omega^2 r^2 \phi + g ( N - 1 ) |\phi|^2 \phi = \epsilon \phi \,
\end{equation}

The solution $\phi$ must be normalized by the condition $\int |\phi|^2 \mathrm{d}^3 \vec{r} = 1$. $\epsilon$ is the chemical potential (or orbital energy) of the system, and it is related to the many-body energy $E$ by the relation $E = N \epsilon - \frac{1}{2} g N ( N - 1 ) \int |\phi|^4 \mathrm{d}^3 \vec{r}$. For the price of reducing the many-body problem to an equation for a single particle, the interaction is now represented by a nonlinear mean-field term. The above equation gives the ground-state properties of the system, and the lowest few excitations are usually treated in terms of Bogoliubov modes \cite{PhysRevA.55.1147}.

Here, an alternative for the many-body problem is presented. From now on, a dimensionless system of units is adopted, where length is in units of the trap oscillator length $l_t = \sqrt{\frac{\hbar}{m \omega}}$ and energy is in units of $\hbar \omega$. The hyperradius is defined as $R = \sqrt{ \frac{1}{N} \displaystyle\sum_{i=1}^N r_i^2 }$. The $3N$ Cartesian coordinate system is recast in a hyperspherical coordinate system $( R, \Omega )$, with $3N-1$ hyperangles $\Omega$ describing the internal particle configurations of a fixed hypersphere of radius $\sqrt{N} R$. The choice of coefficient $\frac{1}{N}$ in the definition of the hyperradius is convenient, allowing an intuitive meaning of $R$ as the root-mean-square of the individual particle distance from the center of the trap, giving an overall size of the atomic gas.

The Laplacian and the many-body Hamiltonian take the following form:

\begin{align}
    \displaystyle\sum_{i=1}^N \nabla_i^2 &= \frac{1}{N} \left( \frac{1}{R^{3N-1}} \frac{\partial}{\partial R} \left( R^{3N-1} \frac{\partial}{\partial R} \right) - \frac{\Lambda^2}{R^2} \right) \\
    H &= - \frac{1}{2N} \frac{1}{R^\frac{3N-1}{2}} \frac{\partial^2}{\partial R^2} R^\frac{3N-1}{2} + H_A \\
    H_A &= \frac{1}{2N} \left( \frac{(3N-1)(3N-3)}{4 R^2} + \frac{\Lambda^2}{R^2} \right) + \frac{1}{2} N R^2 \nonumber \\
    &+ 4 \pi a_s \displaystyle\sum_{i<j} \delta ( \vec{r_i} - \vec{r_j} )
\end{align}

The external trap potential takes a simple form in the hyperspherical coordinates, depending only on the hyperradius. The kinetic energy operator is written in terms of a simple second-derivative in $R$, a repulsive centrifugal term proportional to $\frac{1}{R^2}$, and contributions from a grand angular-momentum operator $\Lambda^2$ in terms of $\Omega$. There exist \cite{avery2012hyperspherical} a complete, orthonormal basis of hyperspherical harmonics $Y_{\lambda,\mu} ( \Omega ) $ that obey the eigenvalue relation $ \Lambda^2 Y_{\lambda,\mu} = \lambda ( \lambda + 3N - 2 ) Y_{\lambda,\mu} $, with integer $ \lambda = 0, 1, 2, \ldots$ and quantum numbers $\mu$ distinguishing the different degenerate states. The adiabatic formulation makes a quasi-separable ansatz for the energy eigenfunctions, based on the separation of the Hamiltonian into two parts: a derivative term in $R$ and an adiabatic $H_A$ for which $R$ is a fixed parameter.

Let the bra-ket notation denote an integration in the hyperangles with a fixed value of $R$, $\int \mathrm{d}\Omega$. It would be highly challenging, if not impossible, to accomplish a full, exact diagonalization of $H_A$ at various fixed values of $R$ for a general many-body problem, although it is now routinely done for three or four particles. A reasonable alternative is as follows. Write an \textit{ansatz} wavefunction $\Psi = \frac{F(R)}{R^{\frac{3N-1}{2}}} \frac{B(R,\Omega)}{\sqrt{C(R)}}$, where $C(R) = \braket{B|B}$ and $B(R,\Omega)$ is some chosen trial function (assume real). Projecting $B$ onto $H\Psi=E\Psi$ gives an effective \textit{linear} Schr{\"o}dinger equation in $R$ (let $'$ denote $\frac{\partial}{\partial R}$):

\begin{align}
    -\frac{1}{2N} F''(R) &- \frac{Q(R)}{2N} F(R) + \frac{\braket{B|H_A|B}}{C} F(R) = E F(R) \label{Feqn} \\
    Q(R) &= \frac{\braket{B|B''}}{C} + \frac{1}{4} \left( \frac{C'}{C} \right)^2 - \frac{1}{2} \frac{C''}{C}
\end{align}

This is a variational formulation of the adiabatic hyperspherical approach that is widely studied in few-body problems \cite{RevModPhys.89.035006}. In principle, there exists a set of adiabatic channel functions $\Phi_\mu ( R, \Omega )$ that diagonalize $H_A$ at each $R$, $H_A \Phi_\mu ( R, \Omega ) = U_\mu ( R ) \Phi_\mu ( R, \Omega )$, $\mu = 0, 1, \ldots$ with $\braket{\Phi_\mu|\Phi_\nu} = \delta_{\mu \nu}$, so that $\frac{\braket{B|H_A|B}}{C} \geq U_0 ( R )$. $Q(R)$ is a non-adiabatic correction to this approach: the smaller $Q$ is, the more accurate is the adiabatic treatment, provided the solutions are accurate approximations to the true eigenfunctions of the fixed-$R$ Hamiltonian.

The K-Harmonic approximation \cite{PhysRevA.58.584} takes $B(R,\Omega)$ to be merely the lowest hyperspherical harmonics $Y_{0,0} ( \Omega )$, which is in fact just a constant. It is worth noting that for \textit{non-interacting} bosons, the ground state is simply a product of gaussians, $\displaystyle\prod_{i=1}^N e^{-\frac{1}{2}r_i^2} = e^{-\frac{1}{2} N R^2}$, also with no dependence on the hyperangles. For that crude approximation to the adiabatic eigenfunction, $Q(R) = 0$ and $U(R) = \frac{\braket{B|H_A|B}}{C}$ takes a simple analytic form. But bearing in mind that the mean-field solutions for typical laboratory conditions deviate largely from a gaussian, a generalized variational basis set is now chosen with the form of \textit{single-orbital} $B(R,\Omega) = \displaystyle\prod_{i=1}^N \phi ( \vec{r_i} )$. Notice that only the product $a_s ( N - 1 )$ matters in determining the shape of the solution of the GP equation. In calculating the adiabatic hyperspherical potential for a given set of $N$ and $a_s$, one may substitute $N_0 \neq N$ for the GP equation and obtain a corresponding set of solutions $\phi$ to be used as input variational ansatz (or, equivalently, fixed $N$ and different values of $a_s$). Even an arbitrary $\phi$ that has nothing to do with the GP equation is within reach here. $\phi( \vec{r_i} ) = \phi ( r_i )$ is chosen to be a real function of zero angular momentum, properly normalized in all space, and the many-body wavefunction is assumed to be exchange symmetric in the simple product form. A more generalized wavefunction of the form $\Hat{\mathcal{S}} [ \phi_1 ( \vec{r_1} ) \ldots \phi_N ( \vec{r_N} ) ]$, with $\Hat{\mathcal{S}}$ a symmetrization operator, is beyond the scope of this paper. For such a single-orbital trial wavefunction, only certain hyperangular integrals need to be performed (see Appendix) without requiring a diagonalization procedure.

\subsection{Multi-Orbital Extensions}

Next, deviations from a simple product form for the trial wavefunction are considered. This is accomplished by choosing the following structure (given some $n > 1$) of \textit{multi-orbital} $B( R, \Omega ) = \displaystyle\sum_{\mu=1}^n D_\mu ( R ) \frac{B_\mu ( R, \Omega )}{\sqrt{C_\mu ( R )}}$, where $B_\mu(R,\Omega) = \displaystyle\prod_{i=1}^N \phi_\mu ( \vec{r_i} )$ and $D_\mu$ are some expansion coefficients. $\vec{D} = ( D_1, \ldots, D_n )$ is a corresponding vector denoting the particular linear combination of $B_\mu$'s. Denote $C_{\mu \nu} = \braket{B_\mu | B_\nu}$ and $C_{\mu \mu} = C_\mu$, and each $C_\mu$ is different from the overall hyperangular normalization $C$. Such an ansatz is similar in spirit to the configuration-interaction (CI) method commonly used in quantum chemistry \cite{PhysRev.122.1826}. The idea also bears some similarity to the method of eigenvector continuation \cite{PhysRevLett.121.032501}, where eigenstates of a set of model Hamiltonians are used to approximately diagonalize a different Hamiltonian.

What are needed for this treatment are the following five $n$ by $n$ matrices, \underline{$O$}, \underline{$H$}, \underline{$P$}, \underline{$Q$} (not to be confused with the quantity $Q(R)$), and \underline{$P^2$} (not \underline{$P$} x \underline{$P$}), whose matrix elements are as follows:

\begin{align}
    O_{\mu \nu} &= \braket{\frac{B_\mu}{\sqrt{C_\mu}} | \frac{B_\nu}{\sqrt{C_\nu}} } = \frac{C_{\mu \nu}}{ \sqrt{ C_\mu C_\nu } } \\
    H_{\mu \nu} &= \braket{\frac{B_\mu}{\sqrt{C_\mu}} | H_A | \frac{B_\nu}{\sqrt{C_\nu}} } = \frac{ \braket{B_\mu | H_A | B_\nu} }{ \sqrt{ C_\mu C_\nu } } \\
    P_{\mu \nu} &= \braket{\frac{B_\mu}{\sqrt{C_\mu}} | \frac{\partial}{\partial R} | \frac{B_\nu}{\sqrt{C_\nu}} } = \frac{\braket{B_\mu | B'_\nu}}{\sqrt{C_\mu C_\nu}} - \frac{C'_\nu}{2 C_\nu} O_{\mu \nu} \\
    Q_{\mu \nu} &= \braket{\frac{B_\mu}{\sqrt{C_\mu}} | \frac{\partial^2}{\partial R^2} | \frac{B_\nu}{\sqrt{C_\nu}} } \nonumber = \frac{\braket{B_\mu | B''_\nu}}{\sqrt{C_\mu C_\nu}} \nonumber \\
    &- \frac{C'_\nu}{C_\nu} \frac{\braket{B_\mu | B'_\nu}}{\sqrt{C_\mu C_\nu}} + \left( \frac{3}{4} \left( \frac{C'_\nu}{C_\nu} \right)^2 - \frac{C''_\nu}{2 C_\nu} \right) O_{\mu \nu} \\
    P^2_{\mu \nu} &= \braket{\frac{\partial}{\partial R} \left( \frac{B_\mu}{\sqrt{C_\mu}} \right) | \frac{\partial}{\partial R} \left( \frac{B_\nu}{\sqrt{C_\nu}} \right) } = \frac{\braket{B'_\mu | B'_\nu}}{\sqrt{C_\mu C_\nu}} \nonumber \\
    &- \frac{C'_\mu}{2 C_\mu} \frac{\braket{B_\mu | B'_\nu}}{\sqrt{C_\mu C_\nu}} - \frac{C'_\nu}{2 C_\nu} \frac{\braket{B_\nu | B'_\mu}}{\sqrt{C_\mu C_\nu}} + \frac{C'_\mu C'_\nu}{4 C_\mu C_\nu} O_{\mu \nu}
\end{align}

One evaluates the matrix elements of the adiabatic Hamiltonian as follows, using the exchange-symmetric properties of the basis:

\begin{align}
    \braket{B_\mu | H_A | B_\nu} &= \frac{\braket{B_\mu | \Lambda^2 | B_\nu}}{2 N R^2} \nonumber \\
    &+ \left( \frac{(3N-1)(3N-3)}{8 N R^2} + \frac{1}{2} N R^2 \right) C_{\mu \nu} \nonumber \\
    &+ 4 \pi a_s \left( \frac{N(N-1)}{2} \right) \braket{ B_\mu | \delta ( \vec{r_2} - \vec{r_1} ) | B_\nu } \label{adiaH_matrix_element} \\
    \braket{B_\mu | \Lambda^2 | B_\nu} &= R^2 \Big( \braket{B_\mu | B''_\nu } + \frac{3N-1}{R} \braket{B_\mu | B'_\nu} \nonumber \\
    &- N^2 \braket{B_\mu | \nabla_1^2 | B_\nu } \Big)
\end{align}

The matrices \underline{$O$}, \underline{$H$}, and \underline{$P^2$} are symmetric, but \underline{$P$} and \underline{$Q$} are not. \underline{$O$} is the overlap matrix, with $O_{\mu \mu} = 1$, but the off-diagonal elements do not generally vanish since this is a non-orthogonal basis. In principle \underline{$O$}$'$ = \underline{$P$} + \underline{$P$}$^T$ and \underline{$O$}$''$ = \underline{$Q$} + \underline{$Q$}$^T$ + 2 \underline{$P^2$}, with $P_{\mu \mu} = 0$ and $Q_{\mu \mu} = - P^2_{\mu \mu}$.

The overall normalization integral is $C(R) = \braket{B|B} = \vec{D}^T \underline{O} \vec{D} $, and $\braket{B|H_A|B} = \vec{D}^T \underline{H} \vec{D}$. Next, the generalized eigenvalue problem is solved, namely $ \underline{H} \vec{D} = U(R) \underline{O} \vec{D}$, at each fixed value of $R$, and then $\frac{\braket{B|H_A|B}}{C} = U$ for the hyperradial equation of $F(R)$. 

Also, one writes the radial derivative of normalization as $C' = 2 \vec{D}^T \underline{O} \vec{D}' + \vec{D}^T \underline{O}' \vec{D}$, $C'' = 2 \vec{D}'^T \underline{O} \vec{D}' + 2 \vec{D}^T \underline{O} \vec{D}'' + 4 \vec{D}^T \underline{O}' \vec{D}' + \vec{D}^T \underline{O}'' \vec{D}$, and $\braket{B|B''} = \vec{D}^T \underline{Q} \vec{D} + 2 \vec{D}^T \underline{P} \vec{D}' + \vec{D}^T \underline{O} \vec{D}''$. First eliminate the term $\vec{D}^T \underline{O} \vec{D}''$ in $Q(R)$, and then impose the normalization condition $C(R) = \vec{D}^T \underline{O} \vec{D} = 1$ at each $R$ for the eigenvector $\vec{D}$. The non-adiabatic correction $Q(R)$ for the corresponding $B(R,\Omega)$ can then be calculated as: 

\begin{align}
    Q(R) &= - \vec{D}'^T \underline{O} \vec{D}' - 2 \vec{D}^T \underline{P}^T \vec{D}' \nonumber \\
    &+ \vec{D}^T \left( \frac{1}{2} ( \underline{Q} - \underline{Q}^T ) - \underline{P^2} \right) \vec{D}
\end{align}

Finally, to compute $\vec{D}'$, first differentiate both sides of the generalized eigenvalue equation \cite{jwangthesis}. Next, solve the system $( \underline{H} - U \underline{O} ) \vec{Y} = ( - \underline{H}' + U' \underline{O} + U \underline{O}' ) \vec{D}$, where $\vec{Y} = \vec{D}' + c_D \vec{D}$ for some initially unknown coefficient $c_D$. Using the condition $\vec{D}^T \underline{O} \vec{D} = 1$, then $c_D = \vec{D}^T \underline{O} \vec{Y} + \frac{1}{2} \vec{D}^T \underline{O}' \vec{D}$. This type of method is also applied to find the derivative of any other vector as arises below.

\subsection{Treatment of Linear Dependence}

Unfortunately, given a set of chosen orbitals $\phi_\mu$'s, approximate linear dependence can arise between the different variational basis functions $B_\mu (R,\Omega)$, which can result in instability that can in turn produce unphysical generalized eigenvalues $U(R)$. Linear dependence issues also arise in Ref. \cite{PhysRevA.80.022504}, for instance, given a particular choice of correlated gaussian basis set used to compute few-body hyperspherical potential curves. To remedy this should such pathologies arise, the eigenvalue problem can be stabilized by a common procedure as follows. First diagonalize the overlap matrix at each $R$: $\underline{O} \vec{X_l} = o_l \vec{X_l}$. Sort the eigenvalues such that $o_1 \geq \ldots \geq o_n$, and define the corresponding orthogonal eigenvector matrix $\underline{X} = \left( \vec{X_1}, \ldots, \vec{X_n} \right)$.

Consider now the representation where $\underline{\Tilde{O}} = \underline{X}^T \underline{O} \underline{X}$ is diagonal. With $\underline{\Tilde{H}} = \underline{X}^T \underline{H} \underline{X}$ and $\vec{\Tilde{D}} = \underline{X}^T \vec{D}$, the generalized eigenvalue problem in the new representation is $\underline{\Tilde{H}} \vec{\Tilde{D}} = U(R) \underline{\Tilde{O}} \vec{\Tilde{D}}$. So far, everything is equivalent. However, empirically speaking, if at least one eigenvalue of $\underline{O}$ is smaller than some threshold value (typically $10^{-4}$), the eigenvalues $U$ quickly and unphysically collapse towards $-\infty$.

The fix is to choose some $c < n$ and define a submatrix $\underline{X_c}$ of $\underline{X}$: $\underline{X_c} = \left( \vec{X_1}, \ldots, \vec{X_c} \right)$. Be aware that $\underline{X_c}^T \underline{X_c} = \underline{1_c}$, but $\underline{X_c} \underline{X_c}^T \neq \underline{1_n}$. The point is to systematically reduce the dimension of the basis set, so that $B(R,\Omega)$ is composed of $c$, not $n$, basis functions, each of which is a suitable linear combination of $B_\mu$'s. Those linear combinations of $B_\mu$'s with very small corresponding eigenvalues of $\underline{O}$ are nearly $0$ with mostly cancellations amongst each other; they are irrelevant and discarded. For concreteness, define $\underline{\Tilde{O_c}} = \underline{X_c}^T \underline{O} \underline{X_c}$ and $\underline{\Tilde{H_c}} = \underline{X_c}^T \underline{H} \underline{X_c}$, and solve instead the problem $\underline{\Tilde{H_c}} \vec{\Tilde{D}} = \Tilde{U}(R) \underline{\Tilde{O_c}} \vec{\Tilde{D}}$. This is referred to as the \textit{reduced} representation throughout the paper. This results in $c$ eigenvalues $\Tilde{U}$ that are different from the $n$ eigenvalues $U$ in the \textit{primitive} (original) representation, obeying the Hylleraas-Undheim theorem \cite{PhysRevE.84.046705} as the dimension is reduced one-by-one. These $\Tilde{U}$ are then taken for $\frac{\braket{B|H_A|B}}{C}$ in the variational formulation.

The normalization condition for the eigenvectors is now $\vec{\Tilde{D}}^T \underline{\Tilde{O_c}} \vec{\Tilde{D}} = 1$. After a lengthy simplification, the corresponding expression for $Q(R)$ is $Q(R) = - \vec{Z}'^T \underline{O} \vec{Z}' - 2 \vec{Z}^T \underline{P}^T \vec{Z}' + \vec{Z}^T \left( \frac{1}{2} ( \underline{Q} - \underline{Q}^T ) - \underline{P^2} \right) \vec{Z}$, where $\vec{Z} = \underline{X_c} \vec{\Tilde{D}}$. Of course, if $c = n$ and no linear combination of $B_\mu$'s has been eliminated, this is completely equivalent to the expression for $Q(R)$ in the primitive representation.

\section{Results}

Prototypical examples are first considered for repulsive atoms, setting $a_s = 0.01 \, l_t$ for values of $N$ up to $10^4$, to a regime where the Thomas-Fermi approximation should be valid. Also, the conditions of Ref. \cite{PhysRevLett.78.985} are simulated, where \textsuperscript{7}Li has negative scattering length $a_s = -27.3 \, a_0$ and the trap is almost spherically symmetric with oscillator length $l_t = 3.157 \, \mu m$, thus $a_s = -4.577 \times 10^{-4} \, l_t$. $N_c = 1257$ is the largest particle number for which Eq.\ref{GPeqn} has a solution, while the K-Harmonic method predicts $N_c = 1465$ to be the largest $N$ that supports a local minimum in the adiabatic potential.

\begin{figure}[ht]
    \subfloat[$U(R) - \frac{Q(R)}{2 N}$]{\label{apos_singleGPorb_manyN_U}\includegraphics[width=\columnwidth]{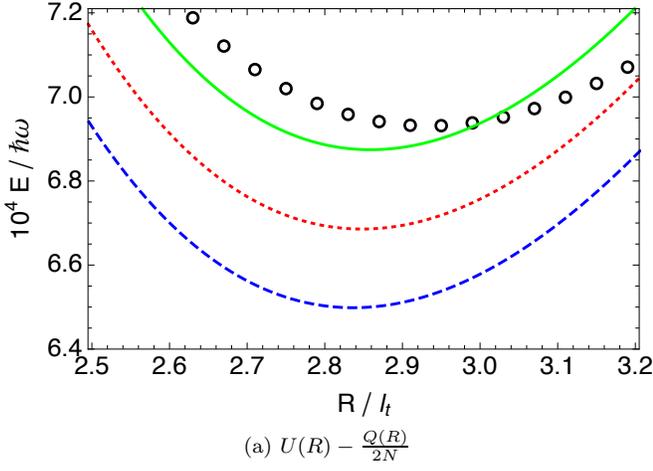}} \\
    \subfloat[$E_0$ and potential minima]{\label{apos_singleGPorb_manyN_minE}\includegraphics[width=\columnwidth]{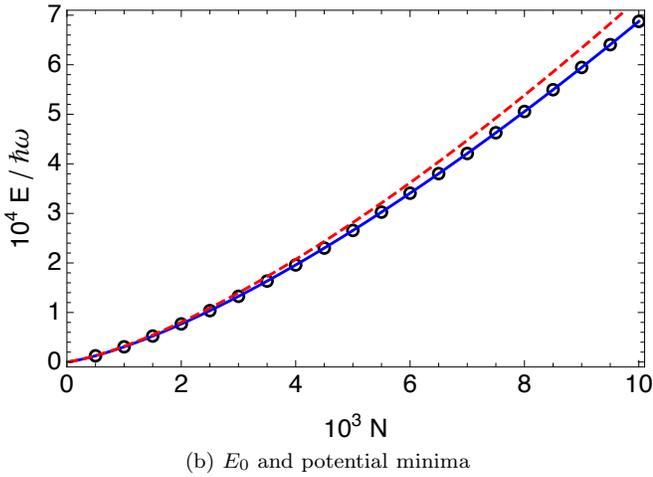}}
    \caption{(Color online)  (a) Adiabatic potential with $Q$ for $a_s = 0.01 \, l_t$. The dashed, dotted, and solid curves come from single $\phi^{GP}$ for $N$ = 9600, 9800, and 10000, respectively. The circles are the K-Harmonic $U$ for $N = 9600$. (b) The circles are the mean-field energy $E_0(N)$. The solid and dashed curves are the minima of $U - \frac{Q}{2N}$ from $\phi^{GP}$ and from the K-Harmonic approximation, respectively.} 
    \label{apos_singleGPorb_manyN}
\end{figure}

As for the types of orbitals used, $\phi_{N_0, a_s}$ is the solution of Eq. \ref{GPeqn} with interaction term $g ( N_0 - 1 ) |\phi|^2 \phi$, if $N_0$ is different from the physical $N$. $\phi^{GP} = \phi_{N,a_s} $ is the correct Gross-Pitaevskii solution, if it exists. Furthermore, for $a_s < 0$, one also considers for given $l>0$:

\begin{equation}
    \phi^{s}_l = \frac{1}{\sqrt{4 \pi}} \sqrt{\frac{12}{\pi^2 l^3}} \, \mathrm{sech} \left( \frac{r}{l} \right)
\end{equation}

The hyperbolic secant is the well-known bright soliton solution to the one-dimensional GP equation \cite{Khaykovich1290}, where the attractive nonlinear term supports a self-bound droplet in the absence of axial trap, and it reasonably approximates $\phi^{GP}$ even for spherically symmetric systems. $\phi^s_l$ is particularly useful for modeling situations where Eq. \ref{GPeqn} has no solution. 

\begin{figure}[t]
    \subfloat[$U(R) - \frac{Q(R)}{2 N}$]{\label{aneg_singleGPorb_manyN_U}\includegraphics[width=\columnwidth]{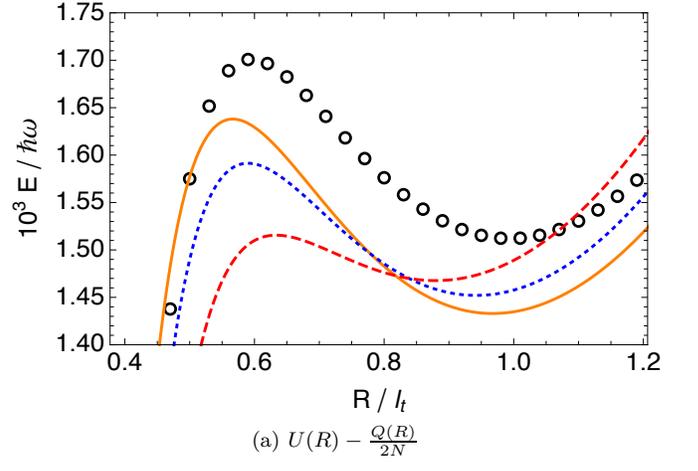}} \\
    \subfloat[$E_0$ and potential minima, maxima]{\label{aneg_singleGPorb_manyN_minE}\includegraphics[width=\columnwidth]{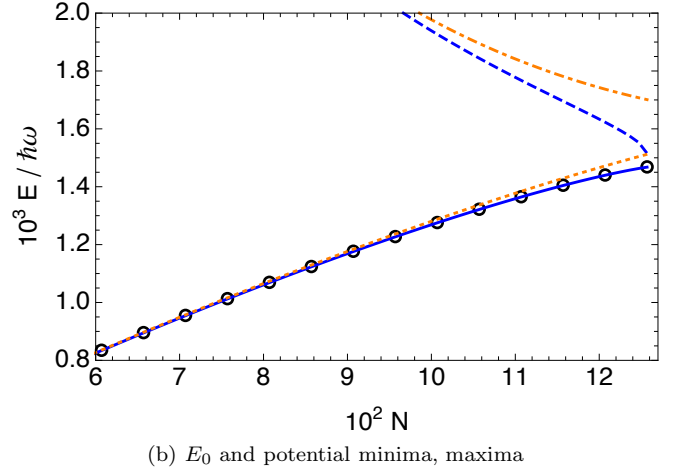}}
    \caption{(Color online) (a) Adiabatic potential with $Q$ for $a_s = -4.577 \times 10^{-4} \, l_t$. The solid, dotted, and dashed curves come from single $\phi^{GP}$ for $N$ = 1197, 1227, and 1257, respectively. The circles are the K-Harmonic potential $U$ for $N = 1257$. (b) The circles are the mean-field $E_0(N)$. The solid and dashed curves are the minima and maxima, respectively, of $U - \frac{Q}{2N}$ from $\phi^{GP}$. The dotted and dash-dotted curves are the minima and maxima, respectively, of the K-Harmonic $U$. }
    \label{aneg_singleGPorb_manyN}
\end{figure}

Fig. \ref{apos_singleGPorb_manyN_U} shows single-$\phi^{GP}$ potential energy curves for positive $a_s$ of varying $N$, and Fig. \ref{aneg_singleGPorb_manyN_U} shows potentials for negative $a_s$. The $\phi^{GP}$ minimizes the minimum of the potential significantly better than the K-Harmonic model does. As $N$ increases, the value of the minimum steadily increases as well; the location of the minimum pushes outward for $a_s > 0$ and draws inward for $a_s < 0$, consistent with the trend in the shapes of the mean-field wavefunction. Furthermore, Figs. \ref{apos_singleGPorb_manyN_minE} and \ref{aneg_singleGPorb_manyN_minE} show that the total mean-field ground-state energy $E_0$ of the system (from Eq. \ref{GPeqn}) is consistent with the value of the potential minimum; $E_0$ is only slightly higher than the minimum, which accounts for the zero-point energy of $F(R)$. In Fig. \ref{aneg_singleGPorb_manyN_U}, the barrier that temporarily protects the metastable condensate from macroscopic collapse decreases as $N$ increases. There is a great difference in barrier height between the K-Harmonic ($\Delta E = 188.99 \, \hbar \omega$) and single-$\phi^{GP}$ ($\Delta E = 47.87 \, \hbar \omega$) models for $N = 1257$, consistent with the over-estimation of $N_c$ by a gaussian orbital. Surprisingly, $\phi^{GP}$ still admits a significant barrier for $N = 1257$, even though one would expect the barrier to disappear since the GP equation has no solution for larger $N$.

\begin{figure}[t]
    \subfloat[$a_s = 0.01 \, l_t$]{\label{apos_Bogo_manyN}\includegraphics[width=\columnwidth]{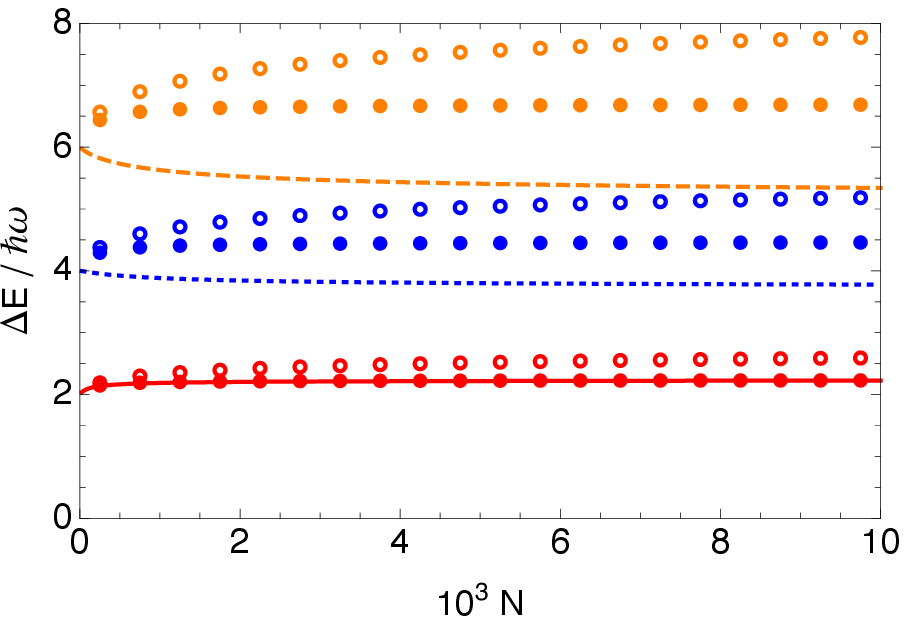}} \\
    \subfloat[$a_s = -4.577 \times 10^{-4} \, l_t$]{\label{aneg_Bogo_manyN}\includegraphics[width=\columnwidth]{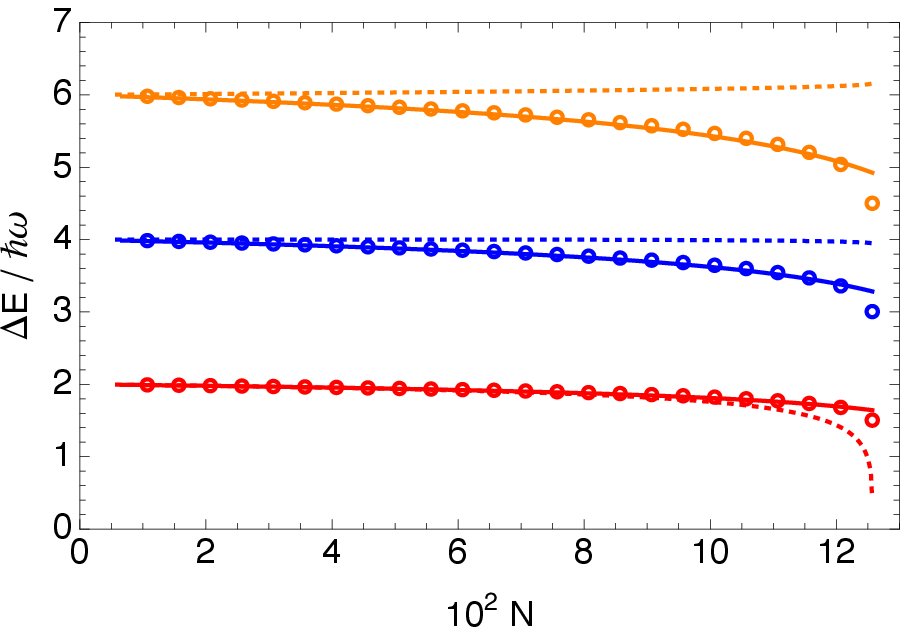}}
    \caption{(Color online) The first $3$ excitation energies $E_i - E_0$ with varying $N$ for positive and negative $a_s$. (a) Solid, dotted, and dashed curves are the Bogoliubov mode frequencies, while solid circles are from the K-Harmonic potentials. Open circles are the results of single $\phi^{GP}$ calculations. (b) Dotted curves are the Bogoliubov predictions, while solid curves are from the K-Harmonic $U$. Circles are from single $\phi^{GP}$ calculations.}
    \label{Bogo_manyN}
\end{figure}

Having checked that the method is reasonably consistent with the mean-field equation in describing the ground state of the many-$N$ bosonic system, the collective excitations are studied next. Given the choice of orbitals with zero angular momentum, one may only hope to reproduce the monopole breathing modes of the spherical system here. Fig. \ref{Bogo_manyN} shows the comparisons between the variational calculations, the K-Harmonic approximation, and the standard Bogoliubov results for the first $3$ breathing modes. In Fig. \ref{apos_Bogo_manyN}, for $a_s > 0$, the Bogoliubov excitation energies transform as $N$ increases from the non-interacting limit of $2 n \hbar \omega$ to the Thomas-Fermi limit \cite{PhysRevLett.77.2360} of $\Delta E_n = \hbar \omega \sqrt{2 n^2 + 3 n}$. The K-Harmonic model agrees well with Bogoliubov theory for the first excited state but gives larger energies for higher states, with a consistent trend of increasing energy with larger $N$. Surprisingly, the single-$\phi^{GP}$ calculations give even greater values for the excitation energies than the K-Harmonic model does. This suggests that the adiabatic potential is too tight; in other words, a simple \textit{ansatz} of $B(R,\Omega) = \displaystyle\prod_{i=1}^N \phi^{GP} ( \vec{r_i} )$ does not accurately capture the more complex many-body correlations and is not sufficient for a variational minimization of $U(R)$ away from the minimum. 

In Fig. \ref{aneg_Bogo_manyN}, for $a_s < 0$, the first Bogoliubov excitation energy tends to rapidly collapse as $N$ approaches $N_c$, while the third and higher excited states are predicted to shoot up. The K-Harmonic model predicts instead that all the excitation energies soften, as long as the corresponding excited states can be supported by the potential barrier. For single-$\phi^{GP}$ calculations, a trend that is the opposite of that for positive $a_s$ is seen: the excitation energies collapse more rapidly than the K-Harmonic model predicts, consistent with the broader curvature of the potential minimum and lower barrier in Fig. \ref{aneg_singleGPorb_manyN_U}. Still, the predicted first excitation energy is higher than the Bogoliubov result.

\begin{figure}[t]
    \includegraphics[width=\columnwidth]{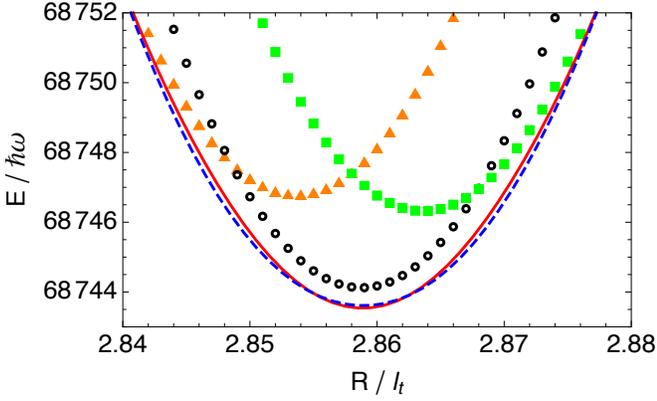}
    \caption{(Color online) Adiabatic potentials with $Q(R)$ included for $10^4$ bosons with $a_s = 10^{-2} \, l_t$. The triangles, circles, and squares are single $\phi_{N_0,a_s}$ results with $N_0 = 9600.04, 10^4$, and $10399.96$. The solid curve is the lowest primitive eigenvalue from coupling $3$ orbitals with $N_0 = 9800.02, 10^4$, and $10199.98$, with no eigenstate of $\underline{O}$ removed. The dashed curve is the lowest reduced eigenvalue from coupling $5$ orbitals with $N_0 = 9600.04, 9800.02, \ldots, 10399.96$, with $1$ eigenstate of $\underline{O}$ removed.}
    \label{apos_multiorb_U}
\end{figure}

These observations provide the key motivation for the coupled multi-orbital CI method, and its results are shown in Fig. \ref{apos_multiorb_U} for $10,000$ repulsive atoms. Within the single-orbital picture, one sees that $\phi^{GP}$ minimizes the ground-state of the system better than any other choice of the orbital, but away from the minimum, other choices of the orbital are superior for a variational minimization of $U(R)$. Therefore, coupling several of these orbitals results in not only an additional lowering of the minimum, but an overall broadening of the curvature as well. For the dashed curve of Fig. \ref{apos_multiorb_U}, the particular choice of coupled $5$ orbitals results in a rapid collapse of the lowest primitive generalized eigenvalue $U(R)$, and hence one eigenstate of \underline{$O$} has been removed. 

On these different results for the adiabatic potential, the hyperradial eigenstates $F(R)$ may now be found, and Table \ref{table_couporb_posa_states} summarizes the results. Compared to the single-$\phi^{GP}$ model, significant lowering of the hyperradial state energies is now observed from multi-orbital calculations, but the excitation frequencies are mostly still higher than what the Bogoliubov theory gives. Actually, for (n,$\delta$) = (5,1) and ($N_{0,1}$,$\Delta N_0$) = ($9600.04,199.98$), the first excitation energy is lower than the corresponding Bogoliubov prediction. It is currently unknown how the variational minimization of $U(R)$ tends to convergence with different $B(R,\Omega)$, or whether it even converges at all. In Ref. \cite{PhysRevA.60.1451}, standard CI calculations (outside the hyperspherical framework) using the pseudopotential have been shown to not converge in the absolute sense. At any rate, the hyperspherical CI method assumes that each term of the wavefunction is a simple product of orbitals, which is a strong restriction on the subspace of Hilbert space that the many-body system occupies, possibly explaining the discrepancies with the Bogoliubov predictions.

%
\begin{table}[t]
\caption{\label{table_couporb_posa_states} List of ground state $E_0$ and excitation $\Delta E$ (units of $\hbar \omega$) from solving Eq.\ref{Feqn} with different $U(R) - \frac{Q(R)}{2N}$, for $N = 10^4$ and $a_s = 10^{-2} \, l_t$. $n$ is the number of coupled orbitals, and $\delta = n - c$ is the number of eigenstates of \underline{$O$} thrown away. The orbitals $\phi_{N_0,a_s}$ are identified by values of $N_0 = N_{0,1} + ( i - 1 ) \Delta N_0$, $i = 1, \ldots, n$.}
\begin{ruledtabular}
\begin{tabular}{|c|c|c|c|c|}
type & (n,$\delta$) & ($N_{0,1}$,$\Delta N_0$) & $E_0$ & $\Delta E$ \\
\hline
K-Har. & (1,0) & n.a. & 73346.48 & 2.23, 4.46, 6.69 \\
\hline
$\phi^{GP}$ & (1,0) & n.a. & 68745.42 & 2.59, 5.19, 7.78 \\
\hline
$\phi_{N_0,a_s}$ & (3,0) & (9800.02,199.98) & 68744.71 & 2.32, 4.60, 6.84 \\
\hline
$\phi_{N_0,a_s}$ & (5,0) & (9500.05,249.975) & 68744.52 & 2.30, 4.55, 6.77 \\
\hline
$\phi_{N_0,a_s}$ & (5,1) & (9600.04,199.98) & 68744.69 & 2.19, 4.43, 6.68 \\
\hline
$\phi_{N_0,a_s}$ & (5,2) & (9800.02,99.99) & 68744.71 & 2.32, 4.59, 6.83 \\
\end{tabular}
\end{ruledtabular}
\end{table}

\begin{figure}[ht]
    \subfloat[$N = 1220$]{\label{aneg_many_singorb_N1220}\includegraphics[width=\columnwidth]{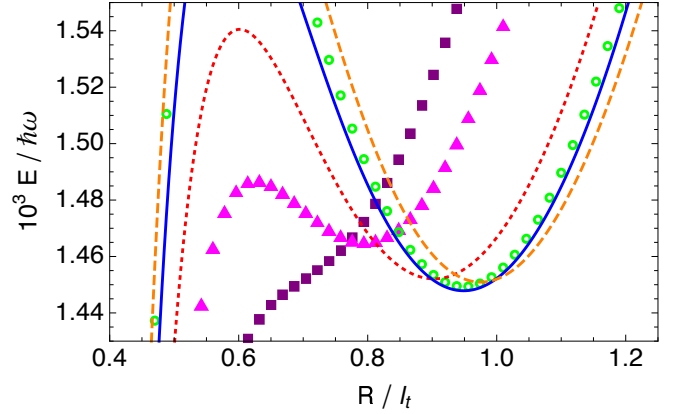}} \\
    \subfloat[$N = 1300$]{\label{aneg_many_singorb_N1300}\includegraphics[width=\columnwidth]{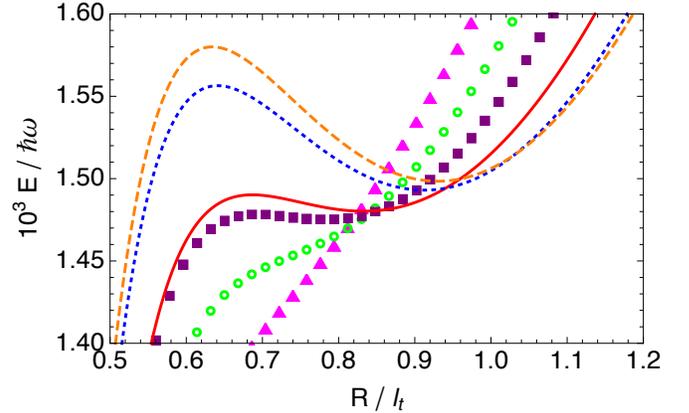}}
    \caption{(Color online) Adiabatic potentials with $Q(R)$ included for $N = 1220$ and $1300$, respectively. Here $a_s = -4.577 \times 10^{-4} \, l_t$, and all are single-orbital results. (a) Dashed, solid, and dotted curves are from $\phi_{N_0, a_s}$ with $N_0 = 1112.41, 1220$, and $1257.1$. Squares, triangles, and circles are from $\phi^{s}_l$ with $l = 0.38, 0.43$, and $0.64$. (b) Dashed, dotted, and solid curves are from $\phi_{N_0, a_s}$ with $N_0 = 1149.27, 1203.19$, and $1257.12$. Triangles, circles, and squares are from $\phi^{s}_l$ with $l = 0.38, 0.43$, and $0.48$.}
    \label{aneg_many_singorb}
\end{figure}

\begin{figure}[ht]
    \includegraphics[width=\columnwidth]{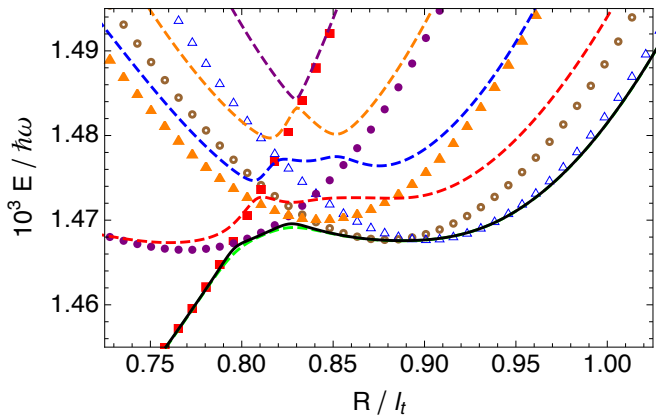}
    \caption{(Color online) Adiabatic potentials for $1257$ bosons with $a_s = -4.577 \times 10^{-4} \, l_t$. Open circles and open triangles are $U$ with $Q$ included from single $\phi_{N_0, a_s}$, with $N_0 = 1257$ and $1251.51$. Squares, filled circles, and filled triangles are $U$ with $Q$ included from single $\phi^{s}_l$ with $l = 0.39, 0.44$, and $0.49$. The $5$ dashed curves are the primitive generalized eigenvalues $U$, \textit{without} $Q$, from coupling the above $5$ orbitals; no eigenstate of \underline{$O$} is removed. The solid curve includes $Q$ for the lowest eigenvalue.}
    \label{aneg_many_singorb_N1257}
\end{figure}

\begin{figure}[ht]
    \includegraphics[width=\columnwidth]{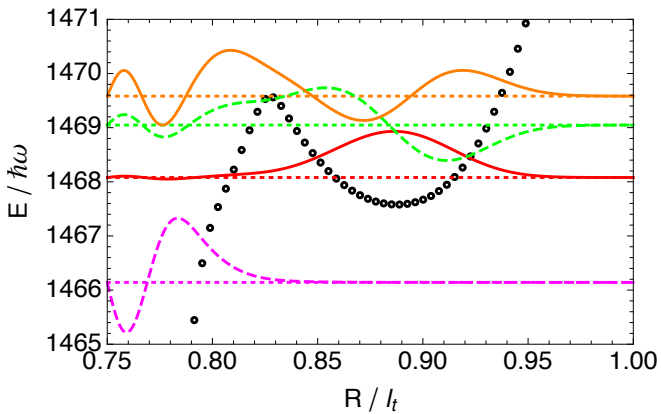}
    \caption{(Color online) Hyperradial eigenstates $F(R)$ for $N = 1257$ and $a_s = -4.577 \times 10^{-4} \, l_t$. The potential, represented by circles, is the solid curve of Fig. \ref{aneg_many_singorb_N1257}. Boundary condition $F(R_c) = 0$ at $R_c = 0.75 \, l_t$ is chosen. Dotted lines denote the eigenenergies; solid and dashed curves are the corresponding wavefunctions $F(R)$ (scaled arbitrarily).}
    \label{aneg_many_singorb_N1257_eigene}
\end{figure}

Now consider $a_s < 0$. Fig. \ref{aneg_many_singorb} shows a family of single-orbital calculations using both Gross-Pitaevskii solutions and bright solitons. In Fig. \ref{aneg_many_singorb_N1220}, again it is seen that $\phi^{GP}$ performs best in minimizing the hyperradial ground state. However, using hyperbolic secant orbitals, one may model situations where the system has been squeezed closer to the origin. Intuitively, one expects that coupling several of the orbitals in Fig. \ref{aneg_many_singorb_N1220} will result in a new potential, which would have a far lower barrier height than the single-$\phi^{GP}$ potential has. For $N = 1300$, $\phi^{GP}$ does not exist, and coupling other available orbitals would now result in a potential with no minimum at all. 

Fig. \ref{aneg_many_singorb_N1257} shows the results of coupling the orbitals for $N = N_c = 1257$. This is the only figure where contributions from non-adiabatic correction $Q(R)$ are explicitly shown, as they were negligible for single-orbital calculations. For the lowest generalized eigenvalue $U(R)$ of interest, contributions from $Q(R)$ are still mostly tiny, except near points where single-orbital potentials cross. Other generalized eigenvalues have large $Q(R)$ near single-orbital potential crossings, exhibiting breakdown of the adiabatic approximation. More importantly, in contrast to the original single-$\phi^{GP}$ result as seen in Fig. \ref{aneg_singleGPorb_manyN_U}, also seen as open circles of Fig. \ref{aneg_many_singorb_N1257}, the coupled-orbital potential now has very small barrier.

Fig. \ref{aneg_many_singorb_N1257_eigene} shows the hyperradial eigenstates from the coupled potential of Fig. \ref{aneg_many_singorb_N1257}. Note that by dimensionality of the pseudopotential, there is an attractive term proportional to $\frac{1}{R^3}$ near the origin in $U(R)$, so unless a small-$R$ cutoff is introduced, the problem of Eq. \ref{Feqn} is ill-defined. An arbitrary boundary condition of $F(R_c) = 0$ at $R_c = 0.75 \, l_t$ was chosen, resulting in collapsed states away from the local minimum, as well as at least one metastable state of energy $E = 1468.08 \, \hbar \omega$ within the minimum. Standard WKB estimate of the macroscopic collapse tunneling lifetime of this metastable state is $0.43 \, s$, while a Siegert pseudostate calculation \cite{PhysRevA.71.032703} approximates the lifetime to be roughly $0.5 \, s$.

The single-orbital method would have predicted the critical particle number $N_c$ where collapse occurs for attractive systems to be far greater than the largest $N$ for which the mean-field equation admits a solution. The coupled multi-orbital method now shows that the criticality of adiabatic hyperspherical potential is consistent with the mean-field equation, but \textit{only} by allowing more many-body correlations than the simple product-symmetric form of the many-body wavefunction assumed in the mean-field equation. Again, it is unknown how the variational potential will converge, especially since now collapse is observed at small $R$ for $a_s < 0$. But the important observation is that the potential barrier has been reduced almost entirely, to the point where it allows only one metastable state for $N = 1257$. Coupling orbitals for $N = 1220$ in Fig. \ref{aneg_many_singorb_N1220}, for example, would lead to a potential that supports many more metastable states. But as $N$ increases toward $1257$, the barrier will decrease and the curvature of local minimum will broaden. One-by-one, each metastable excited state can no longer be supported at some point. This brings into question the validity of the Bogoliubov approximation for treating $a_s < 0$, as it instead predicts that many high-lying excited states not only exist at $N = 1257$, but actually increase in energy compared to the non-interacting limit as seen in Fig. \ref{aneg_Bogo_manyN}.

\section{Conclusion}

New methods for computing the adiabatic hyperspherical potential of many interacting bosons based on the variational principle have been developed, using independent-particle orbitals in connection with the Gross-Pitaevskii equation. Both a very simple product-symmetric form and a configuration-interaction type of variational wavefunction have been investigated. A single-orbital calculation based on the Gross-Pitaevskii solution is found to agree excellently with the mean-field equation itself in computing the ground-state energy. However, systematic differences with the Bogoliubov prediction for monopole excitation energies are observed, and the single-orbital result disagrees with the mean-field equation in predicting the critical number of particles for collapse of attractive system. By coupling several orbitals, a drastic reduction in barrier of the adiabatic potential for $a_s < 0$ is observed, now supporting only one metastable state for $N = N_c$ in agreement with the mean-field prediction.

Several questions and possible future directions remain in describing the dilute quantum gas of many bosons. An immediate possibility would be to generalize the formalism to treat anisotropic traps, in order to investigate quasi-1D and quasi-2D systems. However, more fundamental issues remain unresolved. Convergence properties of the variational potential are unknown, given the singular nature of the pseudopotential. Also, as seen in  Eq. \ref{adiaH_matrix_element}, the interaction term is merely proportional to the scattering length; the variational method presented in this paper is hence inappropriate for describing unitary Bose gas as $|a_s| \to \infty$, for the same reason that the mean-field equation fails at unitarity. Finally, the trial wavefunction does not adequately describe the system as two particles approach each other. The method here using the pseudopotential cannot describe the possibility of clusters of $N-1$ or fewer particles within the $N$-particle system. Therefore, a more pressing problem to be addressed may be to employ realistic finite-range potentials and move beyond the independent-particle approximation, to incorporate the information of two-body correlations into the trial wavefunction for a modest number of particles.

\begin{acknowledgments}
This work was supported in part by the National Science Foundation grant No. PHY-1912350. Hyunwoo Lee thanks P. Giannakeas for helpful conversations.
\end{acknowledgments}

\appendix*
\section{Hyperangular Integration}

This section describes the calculation of the various hyperangular integrals needed. Let spherically symmetric $\phi( \vec{r_i} ) = \frac{1}{\sqrt{4 \pi}} u( r_i )$. For the simplest example, consider $C_{\mu \nu} = \int \mathrm{d}\Omega B_\mu B_\nu = \int \mathrm{d}\Omega \displaystyle\prod_{i=1}^N \phi_\mu ( \vec{r_i} ) \phi_\nu ( \vec{r_i} ) $. At a particular value of the adiabatic parameter $R$, write $ \int \mathrm{d}\Omega = \int \mathrm{d}\Omega \mathrm{d}R' \, \delta ( R - R' ) = N^{-3N/2} R^{-(3N-1)} \int \displaystyle\prod_{i=1}^N \mathrm{d}^3 \vec{r_i}' \delta ( R - R' )$, and use $\delta ( R - R' ) = \frac{N R}{\pi} \int_{-\infty}^{\infty} \mathrm{d}k \, e^{i k N ( R'^2 - R^2 ) }$, following Ref. \cite{PhysRevA.89.012503}. This gives:

\begin{align}
    C_{\mu \nu} ( R ) &= \frac{1}{N^{3N/2} R^{3N-1}} \left( \frac{NR}{\pi}\right) \int_{-\infty}^{\infty} \mathrm{d}k \, e^{-i k N R^2} \nonumber \\
    &\times \left[ \int \mathrm{d}^3 \vec{r'} \, e^{i k {r'}^2 } \phi_\mu ( \vec{r'} ) \phi_\nu ( \vec{r'} ) \right]^N \nonumber \\
    &= \frac{1}{N^{3N/2} R^{3N-1}} \left( \frac{NR}{\pi}\right) \int_{-\infty}^{\infty} \mathrm{d}k \, e^{-i k N R^2} \nonumber \\
    &\times \left[ \int_0^{\infty} \mathrm{d}r' \, r'^2 e^{i k {r'}^2 } u_\mu ( r' ) u_\nu ( r' ) \right]^N
\end{align}

The delta function allows an evaluation of the hyperangular integral in terms of the individual particle coordinates $\vec{r_i}$, which is far easier than trying to express $\phi$ in the hyperspherical coordinate system. For instance, for a term in $H_{\mu \nu}$ coming from the Fermi pseudopotential, dropping primes ($'$) for notational simplicity, one obtains:

\begin{align}
    &\braket{ B_\mu | \delta ( \vec{ r_2 } - \vec{ r_1 } ) | B_\nu } = \left( \frac{1}{4\pi} \right) \frac{1}{N^{3N/2} R^{3N-1}} \left( \frac{NR}{\pi}\right) \nonumber \\
    &\times \int_{-\infty}^{\infty} \mathrm{d}k \, e^{-i k N R^2} \left[ \int_0^{\infty} \mathrm{d}r \, r^2 e^{2 i k r^2 } \left( u_\mu ( r ) u_\nu ( r ) \right)^2 \right] \nonumber \\
    &\times \left[ \int_0^{\infty} \mathrm{d}r \, r^2 e^{i k r^2 } u_\mu ( r ) u_\nu ( r ) \right]^{N-2}
\end{align}

The other integrals to be evaluated are $\braket{ B_\mu | B'_\nu }$, $\braket{ B_\mu | B''_\nu }$, $\braket{ B'_\mu | B'_\nu }$, and $\braket{ B_\mu | \nabla_1^2 | B_\nu }$, where prime denotes $\frac{\partial}{\partial R}$ here.  Using $\frac{\partial}{\partial R} = \displaystyle\sum_{i=1}^N \frac{\partial r_i}{\partial R} \frac{\partial}{\partial r_i} = \displaystyle\sum_{i=1}^N \frac{r_i}{R} \frac{\partial}{\partial r_i}$ and $ \nabla^2 u(r) = \frac{1}{r^2} \frac{\partial}{\partial r} \left( r^2 \frac{\partial u}{\partial r} \right)$, one derives for example:

\begin{align}
    &\braket{ B_\mu | B'_\nu } = \left( \frac{N}{R} \right) \frac{1}{N^{3N/2} R^{3N-1}} \left( \frac{NR}{\pi} \right) \nonumber \\
    &\times \int_{-\infty}^{\infty} \mathrm{d}k \, e^{-i k N R^2} \nonumber \left[ \int_0^{\infty} \mathrm{d}r \, r^3 e^{i k r^2 } u_\mu ( r ) \frac{\partial u_\nu}{\partial r} (r) \right] \nonumber \\
    &\times \left[ \int_0^{\infty} \mathrm{d}r \, r^2 e^{i k r^2 } u_\mu ( r ) u_\nu ( r ) \right]^{N-1}
\end{align}

To note, the following expression can then be derived for the matrix element of $\Lambda^2$, proving that $\braket{B_\mu | \Lambda^2 | B_\nu} = \braket{B_\nu | \Lambda^2 | B_\mu}$:

\begin{align}
    &\braket{B_\mu | \Lambda^2 | B_\nu} = N ( N - 1 )  \frac{1}{N^{3N/2} R^{3N-1}} \left( \frac{NR}{\pi} \right) \nonumber \\
    &\times \int_{-\infty}^{\infty} \mathrm{d}k \, e^{-i k N R^2} \left[ \int_0^{\infty} \mathrm{d}r \, r^2 e^{i k r^2 } u_\mu ( r ) u_\nu ( r ) \right]^{N-2} \nonumber \\
    &\times \Bigg( \left[ \int_0^{\infty} \mathrm{d}r \, r^4 e^{i k r^2 } u_\mu u_\nu \right] \times \left[ \int_0^{\infty} \mathrm{d}r \, r^2 e^{i k r^2 } \frac{\partial u_\mu}{\partial r} \frac{\partial u_\nu}{\partial r} \right] - \nonumber \\
    &\left[ \int_0^{\infty} \mathrm{d}r \, r^3 e^{i k r^2 } u_\mu \frac{\partial u_\nu}{\partial r} \right] \times \left[ \int_0^{\infty} \mathrm{d}r \, r^3 e^{i k r^2 } u_\nu \frac{\partial u_\mu}{\partial r} \right] \Bigg) 
\end{align}

Notice that all the integrals in $k$ above are of the form $\int_{-\infty}^\infty \mathrm{d}k e^{-i k N R^2} g ( k ) \left[ I ( k ) \right]^\beta$. Here $g(k)$ is not being taken to power $N$. Meanwhile $I(k) = \int_0^{\infty} \mathrm{d}r \, r^2 e^{i k r^2 } u_\mu ( r ) u_\nu ( r )$ and $\beta$ is $N$, $N-1$, or $N-2$. Because a factor is being powered to large values of $N$, the integrand oscillates very rapidly on the real line of $k$. The way to proceed is by applying the method of steepest descent \cite{morse1953methods}.

First write $e^{-i k N R^2} \left[ I ( k ) \right]^\beta = e^{ N f ( k ) }$, where $f(k) = - i k R^2 + \frac{\beta}{N} \log I ( k )$. Notice that for $k = i \kappa$, $\kappa \in \mathbb{R}$, assuming that $I(k)$ converges, then $I(k)$ is a real, positive quantity, and hence $f(k)$ is real too. For example, if $u(r) = \frac{2}{\pi^{1/4}} e^{-\frac{r^2}{2}}$, then $I(k) = (1 - i k)^{-3/2}$ if $\mathrm{Im} \, k > -1$. As a function of $R$, there exists a saddle point $k = i \kappa_0$ where $f(k)$ is a minimum on the imaginary axis. By the Cauchy-Riemann equations, with $k = x + i y$, the following conditions hold at $k = i \kappa_0$: $\frac{\partial \mathrm{Re} \, (f)}{\partial x} = 0$, $\frac{\partial \mathrm{Im} \, (f)}{\partial x} = 0$, $\frac{\partial^2 \mathrm{Re} \, (f)}{\partial x^2} < 0$, and $\frac{\partial^2 \mathrm{Im} \, (f)}{\partial x^2} = 0$. Therefore, on the contour $\Gamma$ where $k = x + i \kappa_0$, $x \in ( -\infty, \infty )$, the oscillations in $e^{ N f ( k ) }$ are minimized as the amplitude rapidly decreases away from the saddle point. One may deform the contour and evaluate the resulting smooth integral $\int_{-\infty}^\infty \mathrm{d}k \, g ( k ) e^{ N f ( k ) } = e^{ N f ( i \kappa_0 ) } \int_{\Gamma} \mathrm{d}k \, g ( k ) e^{ N ( f ( k ) - f ( i \kappa_0 ) )}$ by standard numerical quadrature rules.

Now define the following set of even-parity off-centered gaussian fitting functions and their corresponding integral transforms, for some chosen length scale $l$ and distance between the neighboring peaks $r_0$:

\begin{align}
    &\mathcal{B}_n ( l, r_0, r ) = \exp \left( - \left( \frac{r - n r_0}{l} \right)^2 \right) \nonumber \\
    &+ \exp \left( - \left( \frac{r + n r_0}{l} \right)^2 \right) \\
    &\mathcal{B}_m ( \sqrt{2}l, 2r_0, r ) \mathcal{B}_n ( \sqrt{2}l, 2r_0, r ) \nonumber \\
    &= \exp \left( - \frac{(m-n)^2 r_0^2}{l^2} \right) \mathcal{B}_{m+n} ( l, r_0, r ) \nonumber \\
    &+ \exp \left( - \frac{(m+n)^2 r_0^2}{l^2} \right) \mathcal{B}_{|m-n|} ( l, r_0, r ) \\
    &\Tilde{\mathcal{B}}_n ( l, r_0, k ) = \int_0^\infty \mathrm{d}r \, e^{i k r^2} r^2 \mathcal{B}_n ( l, r_0, r ) \nonumber \\
    &= \exp \left( - \frac{(n r_0)^2}{l^2} \left( 1 + \frac{1}{i k l^2 - 1} \right) \right) \nonumber \\
    &\times \Big[ \frac{\sqrt{\pi}}{2} \left( \frac{1}{l^2} - i k \right)^{-3/2} + \sqrt{\pi} \frac{(n r_0)^2}{l^4} \left( \frac{1}{l^2} - i k \right)^{-5/2} \Big] 
\end{align}

To evaluate and analytically continue $I(k)$, one may perform a least-squares fitting approximation with chosen maximum basis index $n_m$ for $\sqrt{ u_\mu ( r ) u_\nu ( r ) }$ ($u_\nu$ if $\mu = \nu$) that is originally expressed in a discrete grid: $ \sqrt{ u_\mu ( r ) u_\nu ( r ) } = \displaystyle\sum_{n=0}^{n_m} \mathcal{C}_n \mathcal{B}_n ( \sqrt{2}l, 2r_0, r )$. Assuming such an expansion is accurate enough, then:

\begin{align}
    &I ( k ) = \int_0^{\infty} \mathrm{d}r \, r^2 e^{i k r^2 } \left( \sqrt{ u_\mu ( r ) u_\nu ( r ) } \right)^2 \nonumber \\
    &= \displaystyle\sum_{m=0}^{n_m} \displaystyle\sum_{n=0}^{n_m} \mathcal{C}_m \mathcal{C}_n \exp \left( - \frac{(m-n)^2 r_0^2}{l^2} \right) \Tilde{\mathcal{B}}_{m+n} ( l, r_0, k ) \nonumber \\
    &+ \displaystyle\sum_{m=0}^{n_m} \displaystyle\sum_{n=0}^{n_m} \mathcal{C}_m \mathcal{C}_n \exp \left( - \frac{(m+n)^2 r_0^2}{l^2} \right) \Tilde{\mathcal{B}}_{|m-n|} ( l, r_0, k )
\end{align}

\begin{figure}[h!]
    \subfloat[Imaginary part of saddle point $k = i \kappa_0$]{\label{appendix_saddle_k}\includegraphics[width=.9\columnwidth]{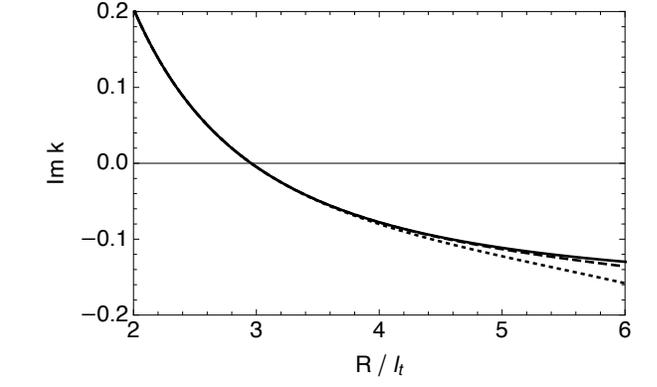}} \\
    \subfloat[Value of $f(k) = - i k R^2 + \log I ( k )$ at $k = i \kappa_0$ ]{\label{appendix_saddle_f}\includegraphics[width=.9\columnwidth]{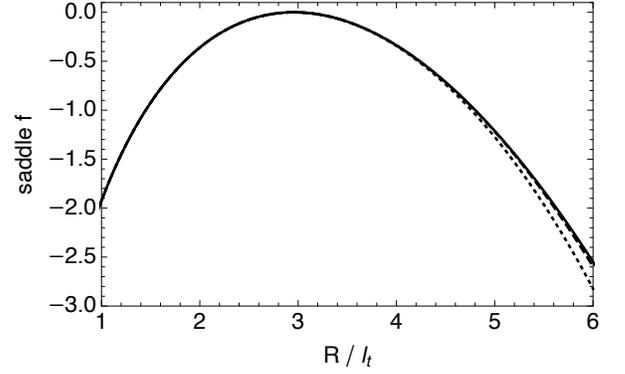}} \\
    \subfloat[Hyperradial logarithmic derivative $\frac{1}{N} \frac{C'}{C}$]{\label{appendix_deriv_c}\includegraphics[width=.9\columnwidth]{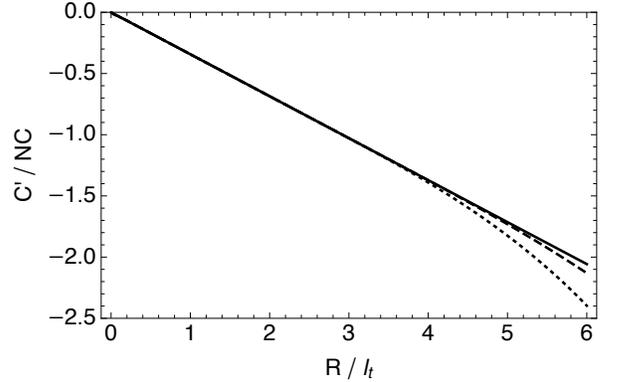}}
    \caption{Benchmark results of single orbital $u = \frac{2}{\pi^{1/4} \alpha^{3/2}} \exp ( -\frac{1}{2} \left( \frac{r}{\alpha} \right)^2 )$ with $\alpha = 2.41529$, for $N = 10^4$ and $a_s = 10^{-2} \, l_t$. (a) and (b) show the saddle point $k = i \kappa_0$ and value of $f(i \kappa_0)$, respectively, with $\beta = N$. (c) shows $\frac{C'}{C}$, divided by $N$. Solid curve is the exact result without least-squares fitting. Dotted curve is the result of approximately fitting $u = \displaystyle\sum_{n=0}^{n_m} \mathcal{C}_n \mathcal{B}_n ( \sqrt{2}l, 2r_0, r )$, with $n_m = 25$, $r_0 = 0.16 \, l_t$, and $l = \sqrt{2} r_0$. Dashed curve is with $n_m = 50$, $r_0 = 0.1 \, l_t$, and $l = \sqrt{2} r_0$.}
    \label{appendix_figures_1}
\end{figure}

Expanding $\sqrt{ u_\mu ( r ) u_\nu ( r ) }$ and taking its square ensures that the resulting approximate $I(k) > 0$ on the imaginary axis of $k$ (where it converges). If, on the other hand, one expands $u_\mu ( r ) u_\nu ( r )$ with $\mu \neq \nu$, then least-squares fitting does not guarantee the positiveness of $I(k)$. The desired saddle point $k = i \kappa_0, \kappa_0 > -\frac{1}{l^2}$, should be found without unphysical difficulties arising from the branch cut of $\log I ( k )$. Furthermore, $l$ should be significantly smaller than the overall size of the orbitals $u_\mu$ and $u_\nu$, not only for good fitting but to ensure that the singularity $k = -\frac{i}{l^2}$ in $\Tilde{\mathcal{B}}_n ( l, r_0, k )$ does not hamper the search for $\kappa_0$.

Similar procedures are employed to express $g(k)$ as well (for $C_{\mu \nu}$, $g = 1$). For instance, for 
$\braket{ B_\mu | B'_\nu }$, $g(k) = \int_0^{\infty} \mathrm{d}r \, r^3 e^{i k r^2 } u_\mu ( r ) \frac{\partial u_\nu}{\partial r} (r)$. Here one approximately expands $\sqrt{ u_\mu \left( -\frac{1}{r} \right) \frac{\partial u_\nu}{\partial r} } = \displaystyle\sum_{n=0}^{n_m} \mathcal{C}_n \mathcal{B}_n ( \sqrt{2}l, 2r_0, r )$ for a different set of coefficients $\mathcal{C}_n$. Then $g(k) = - \int_0^{\infty} \mathrm{d}r \, r^4 e^{i k r^2 } \left( \sqrt{ u_\mu \left( -\frac{1}{r} \right) \frac{\partial u_\nu}{\partial r} } \right)^2$, and an analytic expression for $\int_0^\infty \mathrm{d}r \, e^{i k r^2} r^4 \mathcal{B}_n ( l, r_0, r )$, not $\Tilde{\mathcal{B}}_n ( l, r_0, k )$, is found and used. In the end, only the ratios of quantities such as $\frac{C_{\mu \nu}}{ \sqrt{ C_\mu C_\nu } }$ are needed, so many factors, such as the prefactor $\frac{1}{N^{3N/2} R^{3N-1}} \left( \frac{NR}{\pi}\right)$, cancel out.

To illustrate and benchmark the procedure, consider the conditions of $N = 10^4$ and $a_s = 10^{-2} \, l_t$, and let the single orbital itself be a gaussian, $u_\mu = u_\nu = \frac{2}{\pi^{1/4} \alpha^{3/2}} \exp ( -\frac{1}{2} \left( \frac{r}{\alpha} \right)^2 )$. Use $\alpha = 2.41529$, which variationally minimizes the ground-state energy of the GP equation. Then $I(k)$ and $f(k)$ can be found analytically and the saddle point found numerically without the use of fitting functions, a luxury not afforded to orbitals in general. In fact, all the necessary hyperangular integrals can be done analytically without knowledge of saddle point, giving for example $\frac{C'}{C} = - \frac{2 N R}{\alpha^2}$. Putting such terms together leads to the analytic expression for the K-Harmonic adiabatic potential $U(R)$ in Ref. \cite{PhysRevA.58.584}. Fig. \ref{appendix_figures_1} shows the comparison between exact and fitting function results for the saddle point and $\frac{C'}{C}$. In particular, the dotted curves come from approximating $u$ by 26 fitting functions up to $r = 8 \, l_t$, and dashed curves come from approximation with 51 fitting functions up to $r = 10 \, l_t$.

In the neighborhood of the minimum of $U(R)$, which is at $R = 2.958 \, l_t$ for chosen parameters, excellent agreement between exact and approximate results are seen, as well as convergence in terms of fitting functions. It is seen that $\kappa_0 \to \infty$ as $R \to 0$ and $\kappa_0 \to -\frac{1}{\alpha^2}$ as $R \to \infty$ for the exact result. Interestingly, both $\kappa_0$ and $f(i \kappa_0)$ are nearly $0$ in the vicinity of $R = 2.958 \, l_t$. As is implied by the shape of $f(i \kappa_0)$, plotting $N^{3N/2} R^{3N-1} C$ (which integrates in $R$ to 1) results in an extremely sharp peak at $R = 2.958 \, l_t$, indicating that the system, in a state represented by the gaussian orbital, lies squarely at the minimum of K-Harmonic $U(R)$. Some disagreements between exact and approximate results are observed at small values of $R$, and a more serious deviation is observed at large values of $R$ away from $2.958 \, l_t$. In order to attempt to accurately compute $U(R)$ away from its minimum, more computational effort must be spent to describe the far-lying tail of the orbital with fitting functions. Even then, since the different integrands in $k$ are of the form $e^{ N f( i \kappa_0 )} g(k) e^{ N ( f ( k ) - f( i \kappa_0 ) )}$, serious questions remain regarding the accuracy of the method for large values of $R$. However, since the variational method can only be expected to describe the ground-state and perhaps a few of the lowest-lying breathing modes of the condensate, the method appears satisfactory for the scope of this paper.



\providecommand{\noopsort}[1]{}\providecommand{\singleletter}[1]{#1}%

\end{document}